\documentclass[aps,prl,twocolumn,floatfix,superscriptaddress,longbibliography,notitlepage]{revtex4-2}
\usepackage{braket}
\usepackage{amsmath}
\usepackage{amssymb}
\usepackage[utf8]{inputenc}
\usepackage[T1]{fontenc}
\usepackage{tikz}
\usetikzlibrary{arrows,shapes,calc,matrix}
\usepackage{upgreek} 

\usepackage{changes}
\setdeletedmarkup{\color{blue}\sout{#1}}
\usepackage{cancel}

\usepackage{mathrsfs}
\usepackage[normalem]{ulem}
\usepackage{graphicx, color, graphpap}
\usepackage{enumitem}
\usepackage{amsthm}
\usepackage{multirow}
\usepackage[colorlinks=true,citecolor=blue,linkcolor=blue,urlcolor=blue]{hyperref}
\usepackage{orcidlink}
\usepackage[T1]{fontenc}
\usepackage{verbatim}
\usepackage{mathtools}
\usepackage{titlesec}

\long\def\ca#1\cb{} 

\newcommand{\abs}[2][]{#1| #2 #1|}

\newcommand{\bramatket}[3]{\langle #1 \hspace{1pt} | #2 | \hspace{1pt} #3 \rangle}

\newcommand{\norm}[2][]{#1| \! #1| #2 #1| \! #1|}

\newcommand{\dya}[1]{\ket{#1}\!\bra{#1}}

\newcommand{\Tr}{{\rm Tr}}

\newcommand{\ave}[1]{\langle #1\rangle}               
\renewcommand{\geq}{\geqslant}
\renewcommand{\leq}{\leqslant}

\newcommand*{\id}{\openone}

\newcommand{\la}{\left\langle}
\newcommand{\ra}{\right\rangle}

\newcommand{\bla}{bla\\bla\\bla\\bla\\bla}

\newcommand{\ad}{^\dagger}

\newtheorem{theorem}{Theorem}

\newtheorem*{theorem*}{Theorem}

\begin{document}
\title{Quantum Work Statistics from Entanglement-Assisted Invariance}


\author{Akira Sone
\orcidlink{0000-0003-3539-6140}}
\email{akira.sone@umb.edu}
\affiliation{Department of Physics, University of Massachusetts, Boston, Massachusetts 02125, USA}
\affiliation{The NSF AI Institute for Artificial Intelligence and Fundamental Interactions}

\author{Akram Touil
\orcidlink{0000-0002-2405-7883}}
\affiliation{Theoretical Division, Los Alamos National Laboratory, Los Alamos, NM 87545, USA}

\author{Emery Doucet
\orcidlink{0000-0002-2693-8553}}
\affiliation{Department of Physics, University of Massachusetts, Boston, Massachusetts 02125, USA}

\author{Sebastian Deffner \orcidlink{0000-0003-0504-6932}}
\affiliation{Department of Physics, University of Maryland, Baltimore County, Baltimore, MD 21250, USA}
\affiliation{Quantum Science Institute, University of Maryland, Baltimore County, Baltimore, MD 21250, USA}
\affiliation{National Quantum Laboratory, College Park, MD 20740, USA}

\begin{abstract}
We demonstrate that the one-time measurement scheme is more consistent than the standard two-time measurement scheme with respect to the fundamental symmetry of entanglement, namely entanglement-assisted invariance (envariance). In particular, the work statistics obtained from the one-time measurement scheme arise naturally as a direct consequence of envariance for any driving unitary, without involving an explicit measurement postulate, whereas the analogous identification for the two-time measurement scheme is possible only for a Haar-measure-zero set of unitaries. This envariance-based formulation eliminates the ambiguity in work statistics related to the energetic cost of measurement.  
\end{abstract}

\maketitle

The fluctuation theorems \cite{seifert2012stochastic,esposito2009nonequilibrium}, and most prominently the Jarzynski equality~\cite{Jarzynski97} and the Crooks theorem~\cite{crooks1999entropy}, are the most important results of modern stochastic thermodynamics~\cite{Ortiz2011}. These theorems can be understood as statements of the second law of thermodynamics. At a fundamental level, they express that for small systems, and far from equilibrium, negative excursions of the entropy production can occur, but also that such excursions are exponentially unlikely. Among the various formulations of the fluctuation theorems, the Jarzynski equality \cite{Jarzynski97}
\begin{equation}
\la e^{-\beta W}\ra=e^{-\beta \Delta F}\,,
\end{equation}
might be considered the most influential. Its prominence rests in the fact that the Jarzynski equality provides simple, yet universal means to determine equilibrium free energy differences from nonequilibrium fluctuations. 

It is important to appreciate that fluctuation theorems, such as the Jarzynski equality, were originally only derived for classical systems, for which work is essentially given by ``force times displacement''. Obviously, such a notion of work is inadequate for quantum dynamics, which motivated a plethora of different notions of quantum work. In almost all of these quantum generalizations the role of quantum measurements is accounted for in different ways. This has led to, for instance, the two-time measurement (TTM) scheme~\cite{Tasaki00,Kurchan01,Campisi11,albash2013fluctuation,rastegin2014jarzynski}, work defined with the help of quasiprobability distributions~\cite{Deffner2013EPL,cerisola2023wigner,pei2023exploring,francica2023quasiprobability,gherardini2024quasiprobabilities}, a weak measurement approach~\cite{solinas2017measurement,miller2018leggett,kerremans2022probabilistically}, and the one-time measurement (OTM) formulation~\cite{deffner2016quantum,sone2020quantum,sone2023jarzynski,sone2023exchange,sone2021jarzynski,maeda2023detailed}. 

Interestingly, several of the quantum formulations have been experimentally verified in a wide range of experimental platforms, such as nuclear magnetic resonance (NMR)~\cite{batalhao2014experimental,micadei2021experimental}, trapped-ion~\cite{Smith2018,An15}, photonic~\cite{aguilar2021two,li2025experimental}, nitrogen-vacancy (NV) center~\cite{hernandez2021experimental,hernandez2024interferometry}, and superconducting systems~\cite{Gardas2018SR,solfanelli2021experimental,hahn2023quantum,erdamar2024constraining}. Thus, there does not seem to be immediate experimental evidence whether or not any of these different notions is the ``correct'' approach. Consequently, consensus appears to be lacking as to which notion of quantum work is physically, and more importantly thermodynamically most consistent~\cite{perarnau2017no,deffner2016quantum,richens2016work}.  

While the TTM scheme provides a conceptually clean and experimentally accessible definition of work, it is not immediately obvious how to account for the thermodynamic cost of the required projective measurements~{~\cite{kurt2012quantum,mohammady2021self,sagawa2009minimal,guryanova2020ideal,latune2025thermodynamically,euler}. Progress was made towards answering this question, by interpreting quantum work with the framework of positive-operator-valued measures (POVM)~\cite{roncaglia2014work,cerisola2017using}. In this paradigm, measuring work is understood as \textit{recording} information by coupling the system to an ancillary system (work meter). The interaction generates pure entanglement of the system and memory, and a final projective measurement on the memory effectively implements a generalized measurement on the system. However, the system–memory coupling utilized to encode the work statistics inevitably requires energy exchange; this cost cannot be quantified solely within the TTM approach. These considerations then motivate alternative approaches to defining work statistics that avoid relying on an explicit measurement postulate while still capturing the essential statistical properties of work.

In this Letter, we revisit the work meter models, and consider a different approach to formulating work statistics leveraging entanglement-assisted invariance (or \textit{envariance}),  which is a symmetry-based principle providing a derivation of the Born rule in quantum mechanics~\cite{Zurek2003PRL,Zurek2003RMP,Zurek2005PRA,Zurek2009NatPhys,Zurek2011PRL,zurek2022quantum,Zurek_2025}. As a main result, we demonstrate that the work distribution derived from OTM scheme can be obtained \textit{solely} from envariance, without involving a measurement postulate, for every driving unitary. By contrast, the corresponding identification for the TTM scheme is possible only for a measure-zero set of unitaries.

\paragraph{Preliminaries}

We start by establishing notions and notations. \emph{Envariance} refers to the invariance of a multipartite pure entanglement under local operations on its subsystems. To this end, consider a bipartite system described by the Hilbert space $\mathcal{H}=\mathcal{H}_1\otimes\mathcal{H}_2$. A quantum state in its Schmidt decomposition, 
\begin{align}
\ket{\psi}=\sum_{n}c_n\ket{a_n}_{1}\otimes\ket{b_n}_{2}\,,
\end{align}
where $\ket{a_n}$ and $\ket{b_n}$ form orthogonal bases of $\mathcal{H}_1$ and $\mathcal{H}_2$, is said to be envariant under a local unitary $U_{1}$ on $\mathcal{H}_1$ if its action can be ``undone'' by a corresponding local unitary $U_{2}$ on $\mathcal{H}_2$. In formula,
\begin{align}
\begin{split}
&(U_1\otimes \id_2)\ket{\psi}=\ket{\varphi}\\
&(\id_1\otimes U_2)\ket{\varphi}=\ket{\psi}\,.
\end{split}
\end{align}
Consequently, the combined action of the two transformations leaves the global state invariant
\begin{align}
    \ket{\psi}=(U_1\otimes U_2)\ket{\psi}\,.
\end{align} 
Particularly, the state that is envariant under any local unitary map is referred to as the \textit{maximally envariant state}. This can be illustrated with the following simple example,
\begin{equation*}
    \begin{array}{c}
         \begin{tikzpicture}[>=stealth,baseline,anchor=base,inner sep=0pt]
            \matrix (foil1) [matrix of math nodes,nodes={minimum height=0.1em}] {
                & {\color{blue}\ket{\uparrow}_{1}} &   \otimes   & \ket{\Uparrow}_{2} &  & + &  & {\color{blue}\ket{\downarrow}_{1}} &   \otimes   & \ket{\Downarrow}_{2} &  \xrightarrow{~ {\color{blue}U_{1}}~} 
                {\color{blue}\ket{\downarrow}_{1}}  \otimes  \ket{\Uparrow}_{2}+{\color{blue}\ket{\uparrow}_{1}}  \otimes  \ket{\Downarrow}_{2}\\
            };
            \path[->] ($(foil1-1-2.north)+(0,1ex)$) edge[blue,bend left=45] ($(foil1-1-8.north)+(0,1ex)$);
            \path[<-] ($(foil1-1-2.south)-(0,1ex)$) edge[blue,bend left=-45] ($(foil1-1-8.south)-(0,1ex)$);
        \end{tikzpicture}
        \\[1ex] 
        \begin{tikzpicture}[>=stealth,baseline,anchor=base,inner sep=0pt]
            \matrix (foil2) [matrix of math nodes,nodes={minimum height=0.1em}] {
                & \ket{\downarrow}_{1} &   \otimes   & {\color{red}\ket{\Uparrow}_{2}} &  &+ &  & \ket{\uparrow}_{1} &   \otimes   & {\color{red}\ket{\Downarrow}_{2}} & \xrightarrow{~ {\color{red}U_{2}}~} 
                \ket{\downarrow}_{1}  \otimes  {\color{red}\ket{\Downarrow}_{2}}+\ket{\uparrow}_{1}  \otimes  {\color{red}\ket{\Uparrow}_{2}}\,. \\
            };
            \path[->]  ($(foil2-1-4.north)+(0,1ex)$) edge[red,bend left=45] ($(foil2-1-10.north)+(0,1ex)$);
            \path[<-]  ($(foil2-1-4.south)-(0,1ex)$) edge[red,bend left=-45] ($(foil2-1-10.south)-(0,1ex)$);
        \end{tikzpicture}
    \end{array}
\end{equation*}

Beyond its role in deriving Born rule, envariance has played a central role in advancing our understanding of quantum physics, including quantum-to-classical transitions~\cite{Zurek_2025,touil2024branching}, the foundations of statistical mechanics independent of classical ensembles~\cite{deffner2016foundations,Zurek2018PR,ojha2026statistical}, shortcuts to adiabaticity~\cite{touil2021environment}, and decoherence-free subspace (DFS)~\cite{sone2025no}. In addition, both theoretical analyses and experimental simulations have further elucidated the consequences of envariance~\cite{vermeyden2015experimental,harris2016quantum,deffner2017demonstration}.

In the present context, however, we will leverage envariance as incepted for its original purposes. Namely, we will show that only the probability distribution of the quantum work as constructed from OTM is free of spurious measurements, whereas the TTM approach is fundamentally flawed. 

In previous literature~\cite{deffner2016quantum}, it was already pointed out that a key inconsistency of the TTM approach originates in the fact that work is defined through two projective energy measurements, thereby neglecting the thermodynamic and informational costs associated with performing these measurements. By contrast, in the OTM scheme, work is defined as the expected change in the system's energy conditioned on the outcome of a single initial measurement. This formulation enables an explicit accounting of informational and energetic costs~\cite{deffner2016quantum}, leading to a more stringent formulation of the second law of thermodynamics. 

Beyond these thermodynamic considerations, in the present analysis we offer an additional perspective on the inconsistency of the TTM scheme based on envariance. We demonstrate that, for every driving unitary (except for measure-zero cases), the work distribution obtained from the conventional TTM scheme cannot be justified using envariance, whereas the work distribution associated with the OTM scheme can be naturally derived within the envariance framework for every driving unitary. This distinction can be understood intuitively. The second projective measurement in the TTM scheme destroys the coherence and correlations generated in the quantum meter model. In contrast, in the OTM scheme these coherences and correlations are preserved. Thus, OTM maintains the underlying entanglement symmetry, which then leads to a conceptually clean way to construct the probability distribution of work.

\paragraph{Main results}

Precisely, we will demonstrate that the work distribution obtained from the OTM scheme coincides with the Schmidt coefficients of the entangled state in the work meter model for every driving unitary, whereas the corresponding identification in the TTM scheme is possible only for a Haar-measure-zero set of unitaries. These Schmidt coefficients admit a direct interpretation as Born probabilities that emerge solely from envariance, without requiring any explicit measurement postulate.

To this end, consider a bipartite system $\mathcal{H}=\mathcal{H}_1\otimes\mathcal{H}_2$, {where $\mathcal{H}_1$ is a finite-dimensional system and we treat a continuous degree of freedom, $\mathcal{H}_2$, as a work meter}. Let $H_t$ denote the time-dependent Hamiltonian of the system $\mathcal{H}_1$. 
The unitary time evolution of the system obeys the following the Schr\"odinger equation
\begin{align}
\frac{dU_t}{dt} = -i H_t U_t, \quad U_0 = \id_{1}\,.
\label{eq:evolution}
\end{align}
Let the initial and final Hamiltonian be
\begin{align}
H_0&=\sum_{n}E_n\dya{n}\\
H_{\tau}&=\sum_{m}E_m'\dya{m'}\,,
\end{align}
respectively, where $E_n$ ($E_m'$) and $\ket{n}$ ($\ket{m'}$) are the eigen energies and the corresponding eigenstates of $H_0$ ($H_{\tau}$).

First, we show that, for every driving unitary (except for measure-zero cases), the work distribution in the TTM scheme does not arise as a Schmidt coefficient within the work meter model. Following Ref.~\cite{roncaglia2014work}, we start from the following product state
\begin{align}
    \ket{\psi_{<0}}=\sum_{n}c_n\ket{n}_1\otimes\ket{0}_{2}\,.
\end{align}
We turn on the system-meter interaction described by a unitary $\Gamma_A$, 
\begin{align}
    \Gamma_A = \sum_{n}\dya{n}\otimes e^{iE_n P}\,,
\end{align}
which performs a coherent controlled translation generated by a Hermitian operator $P$ such that
\begin{align}
    e^{i y P}\ket{x}=\ket{x-y}\,.
\end{align}
Note that the operator $P$ satisfies the canonical commutation relation $[X,P]=i$, where $X$ is the Hermitian operator whose eigenstates $\ket{x}$ define the orthogonal basis via $X\ket{x}=x\ket{x}$. 
As a consequence, the interaction acts as a translation on work meter, thereby {recording} the meter states with the energy eigenvalue differences of the system.
In this way, the energy shift is encoded as a \textit{displacement} of the meter. 

After this interaction, we obtain the following entangled state
\begin{align}
\ket{\psi_0} = \Gamma_A\ket{\psi_{<0}}= \sum_{n} c_n 
\ket{n}_{1}   \otimes   \ket{-E_n}_{2}\,,
\label{eq:eq:S+B_entangle_<0}
\end{align}
and the initial reduced state of $\mathcal{H}_1$,
\begin{align}
    \rho_0\equiv \Tr_2\left[\dya{\psi_0}\right]=\sum_{n}\abs{c_n}^2\dya{n}\,.
\label{eq:initial_{1}tate}
\end{align}
We evolve $\ket{\psi_0}$ from $t=0$ to $t=\tau$ under $U_{t} \otimes \id_{2}$, we then have 
\begin{align}
\ket{\psi_{\tau}}=\sum_{n}c_n U_{\tau}\ket{n}_{1} \otimes  \ket{-E_n}_{2}\,,
\label{eq:state_at_tau}
\end{align}
and the final reduced state of $\mathcal{H}_1$,
\begin{align}
    \rho_{\tau}\equiv \Tr_2\left[\dya{\psi_{\tau}}\right]=U_{\tau}\rho_0U_{\tau}\ad\,.
\end{align}

For the TTM scheme, writing the work along a single trajectory as 
\begin{align}
    w_{mn}\equiv E_m'-E_n\,,
\end{align}
we again turn on the system-meter interaction by applying a unitary
\begin{align}
    \Gamma_B = \sum_{m}\dya{m'}\otimes e^{-i E_m' P}\,.
\end{align}
Then, {after discretizing the meter coordinate and representing distinct work value by orthonormal meter states, namely $\braket{w|w'}=\delta_{w,w'}$~\footnote{We employ discretization approximation so that the Kronecker delta, $\delta_{w,w_{mn}}$, can be used to evaluate $\abs{\alpha_w}^2$ in Eq.~\eqref{eq:alpha_TTM}. Without this approximation, the continuous formulation would instead involve the Dirac delta, $\delta_{\rm{D}}(w-w_{mn})$, whose square, $\delta_{\rm{D}}^2(w-w_{mn})$, is not well defined.},} we obtain $\ket{\Psi_{\tau}}=\Gamma_B\ket{\psi_{\tau}}$ as 
\begin{align}
\begin{split}
\ket{\Psi_{\tau}}
&=\sum_{n,m}c_n\bramatket{m'}{U_{\tau}}{n}\ket{m'}_1\otimes\ket{w_{mn}}_2\\
&\approx\sum_{w}\left(\sum_{n,m}c_n\bramatket{m'}{U_{\tau}}{n}\delta_{w,w_{mn}}\ket{m'}_1\right)\otimes\ket{w}_2\\
&=\sum_{w}\alpha_w\ket{\phi_w}_1\otimes\ket{w}_2\,,
\end{split}
\label{eq:TTM_env}
\end{align}
where $\ket{\phi_w}$ is defined as
\begin{align}
\ket{\phi_w}\equiv 
\frac{1}{\alpha_w} \sum_{n,m} c_n\,
\bramatket{m'}{U_{\tau}}{n}\,
\delta_{w,w_{mn}} \ket{m'} \,.
\label{eq:phi_wTTM}
\end{align}
Because $\ket{\phi_w}$ has to be a pure state, using the orthonormal condition 
$\braket{k'|\ell'}=\delta_{k,\ell}$ together with the identity 
$\delta_{w,w_{mn}}^2=\delta_{w,w_{mn}}$, we have 
\begin{align}
  \abs{\alpha_w}^2
  = \sum_{n,m} \abs{c_n}^2
\abs{\bramatket{m'}{U_{\tau}}{n}}^2
    \delta_{w,w_{mn}} \,,
\label{eq:alpha_TTM}
\end{align}
which coincides with the TTM work distribution defined by using Kronecker delta~\cite{kiely2023entropy} when $\abs{c_n}^2$ is a Boltzmann distribution
\begin{align}
    \abs{c_n}^2=\frac{e^{-\beta E_n}}{Z_{\beta}}
\label{eq:boltzmann}
\end{align}
with the canonical partition function $Z_{\beta}=\sum_n e^{-\beta E_n}$, which can be obtained from envariance~\cite{deffner2016foundations}. 

However, the states $\ket{\phi_w}$ do not always form an orthogonal basis of $\mathcal{H}_1$, since in general one can have  $\braket{\phi_w|\phi_{w'}} \neq 0$ for $w \neq w'$. More specifically, from Eq.~\eqref{eq:phi_wTTM}, we obtain
\begin{align}
\braket{\phi_w|\phi_{w'}}
=\delta_{w,w'}+\chi_{w,w'}\,,
\end{align}
where we define
\begin{align}
    \chi_{w,w'} \equiv \sum_{m}\left[\sum_{n}f_{w}^*(m,n)\left(\sum_{k\neq n}f_{w'}(m,k)\right)\right]
\end{align}
with
\begin{align}
    f_{w}(m,n)\equiv\frac{c_n}{\alpha_{w}}\bramatket{m'}{U_{\tau}}{n}\delta_{w,w_{mn}}\,,
\end{align}
which is nonzero when $w=w_{mn}\equiv E_m'-E_n$; therefore, $\chi_{w,w'}$ can take a nonzero value even when $w\neq w'$.

More strongly, if the initial state populates at least two distinct initial energies and the final Hamiltonian has at least two distinct energy levels, the set of $U_{\tau}\in\mathrm{U}(d)$ for which all nonzero-weight TTM work states are mutually orthogonal has Haar measure zero (see Theorem~\ref{thm:TTM_measure_zero} in the Supplementary Information~\cite{supp1}). Thus, at least one pair of nonzero-weight branch states has a nonvanishing overlap, the amplitudes $\alpha_w$ cannot all be interpreted as Schmidt coefficients, and the TTM work distribution does not arise from envariance within this work meter construction.

Next, by contrast, we demonstrate that in the OTM case the work distribution emerges as the Schmidt coefficient when this work meter model is employed. Starting from Eq.~\eqref{eq:state_at_tau}, let the system and the meter interact in a manner different from that of the TTM scheme, which can be represented by a different unitary $\widetilde{\Gamma}_B$, 
\begin{align}
    \widetilde{\Gamma}_B = \sum_{n}U_{\tau}\dya{n}U_{\tau}\ad\otimes e^{-i\widetilde{E}_n P}\,,
\end{align}
where the average energy $\widetilde{E}_n$ conditioned on $E_n$ at $t=\tau$ is defined as 
\begin{align}
\widetilde{E}_n\equiv\bramatket{n}{U_{\tau}\ad H_{\tau}U_{\tau}}{n}\,.
\end{align}
Then, through this interaction, the work meter encodes the conditional work as
\begin{align}
\widetilde{w}_n \equiv \widetilde{E}_n - E_n\,,
\label{eq:conditionalwork}
\end{align}
via 
\begin{align}
    \!\!\!\!\widetilde{\Gamma}_B \big(U_{\tau}\ket{n}_1 \otimes \ket{-E_n}_2\big) = U_{\tau}\ket{n}_1 \otimes \ket{\widetilde{w}_n}_2\,.
\end{align}
Therefore, 
$\ket{\widetilde{\Psi}_{\tau}}=\widetilde{\Gamma}_B \ket{\psi_{\tau}}$ becomes
\begin{align}
\begin{split}
\ket{\widetilde{\Psi}_{\tau}}
&= \sum_{n}c_n U_{\tau}\ket{n}_{1}  \otimes  \ket{\widetilde{w}_n}_{2}\\
&\approx\sum_{w}\left(\sum_{n}c_n\delta_{w,\widetilde{w}_n}U_{\tau}\ket{n}_{1}\right)  \otimes  \ket{w}_{2}\\
&=\sum_w \widetilde{\alpha}_w\ket{\widetilde{\phi}_w}_{1}  \otimes  \ket{w}_{2}\,,
\end{split}
\label{eq:final_joint_state}
\end{align}
where the state $\ket{\widetilde{\phi}_w}$ is given by 
\begin{align}
    \ket{\widetilde{\phi}_w} = \frac{1}{\widetilde{\alpha}_w}\sum_{n}c_n\delta_{w,\widetilde{w}_n}U_{\tau}\ket{n}\,,
\end{align}
and again $\ket{w}$ forms an orthonormal basis of $\mathcal{H}_2$. {Here,  the term $\ket{n}_{1}\otimes\ket{\widetilde{w}_{n}}_{2}$ describes the encoding of the energy transition path,  $E_n \to \widetilde{E}_{n}$. }

Since $\ket{\widetilde{\phi}_w}$ is a pure state, we must have 
\begin{align}
\abs{\widetilde{\alpha}_w}^2=\sum_{n} \abs{c_n}^2 \delta_{w,\widetilde{w}_n}\,,
\label{eq:workstatistics}
\end{align}
which coincides with the OTM work distribution proposed in Ref.~\cite{deffner2016quantum} when $\abs{c_n}^2$ is a Boltzmann distribution in Eq.~\eqref{eq:boltzmann}. More importantly, in contrast to the TTM scenario, the states $\ket{\widetilde{\phi}_w}$ \textit{always} form an orthogonal basis of $\mathcal{H}_1$, which can be demonstrated by the following:
\begin{align}
\begin{split}
\braket{\widetilde{\phi}_w|\widetilde{\phi}_{w'}}
&=\frac{1}{\widetilde{\alpha}_w^*\widetilde{\alpha}_{w'}}\sum_{n,k}c_{n}^{*}c_{k}\delta_{w,\widetilde{w}_n}\delta_{w',\widetilde{w}_{k}}\delta_{n,k}\\
&=\frac{1}{\widetilde{\alpha}_w^*\widetilde{\alpha}_{w'}}\sum_{n}\abs{c_{n}}^2\delta_{w,\widetilde{w}_n}\delta_{w',\widetilde{w}_n}\\
&=\delta_{w,w'}\,.
\end{split}
\end{align}

More formally, the following demonstration makes clear the evidence that $|\widetilde{\alpha}_w|^2$ constitutes the work distribution. Let the Hamiltonian of $\mathcal{H}_2$ be
\begin{align}
    W=\sum_w w\dya{w}\,.
\label{eq:work_hamiltonian}
\end{align}
Then, we can find that 
\begin{align}
\begin{split}
\ave{W}_{\tau}&\equiv\bramatket{\widetilde{\Psi}_{\tau}}{\id_{1} \otimes  W}{\widetilde{\Psi}_{\tau}}\\
&=\sum_w\abs{\widetilde{\alpha}_w}^2w\\
&=\Tr\left[\rho_{\tau}H_{\tau}\right]-\Tr\left[\rho_{0} H_0\right]\,.
\end{split}
\end{align}
which is exactly the average work implemented via $U_{\tau}$. Also, when Eq.~\eqref{eq:boltzmann} holds, the Jarzynski equality for OTM scheme derived in Ref.~\cite{deffner2016quantum} can be recovered as 
\begin{align}
\begin{split}
\ave{e^{-\beta W}}_{\tau}&\equiv\bramatket{\widetilde{\Psi}_{\tau}}{\id_{1} \otimes  e^{-\beta W}}{\widetilde{\Psi}_{\tau}}\\
&=\sum_{w}\abs{\widetilde{\alpha}_w}^2 e^{-\beta w}\\
&=\frac{1}{Z_{\beta}}\sum_{n}e^{-\beta\bramatket{n}{U_{\tau}\ad H_{\tau}U_{\tau}}{n}}\,.
\end{split}
\end{align}
Therefore, $\abs{\widetilde{\alpha}_w}^2$ recovers these statistics as a probability distribution of $w$, and OTM work distribution can be interpreted as a natural consequence of the envariance. This highlights a key advantage of the OTM scheme in the envariance framework as the coefficient $\widetilde{\alpha}_w$ can be directly identified as a Schmidt coefficient for every driving unitary, whereas the corresponding identification in the TTM scheme is available only on a Haar-measure-zero set of driving unitaries. Here, we note that this protocol is physically realizable, for example, in ultracold atomic systems~\cite{cerisola2017using}, where the motional degrees of freedom of the atoms serve as a work meter.

{Lastly, we  note that the TTM work distribution can be derived from envariance if we introduce an additional 
subsystem $\mathcal{H}_{R}$ coupled to the system $\mathcal{H}_{1}$ and the work meter $\mathcal{H}_{2}$ (See Supplemental Information~\cite{supp1} for detailed explanation). Here, $\mathcal{H}_R$ encodes the eigenstate of the initial Hamiltonian into its displacement.  In this case, we apply $\Gamma_{A}=\sum_{n}\dya{n}\otimes e^{-i n P}\otimes e^{iE_{n} P}$ on $\ket{\psi_{<0}}=\sum_{n} c_n 
\ket{n}_{1}   \otimes\ket{0}_{R}\otimes   \ket{-E_{n}}_{2}$. In this way, the initial state becomes $\ket{\psi_0} = \sum_{n} c_n 
\ket{n}_{1}   \otimes\ket{n}_{R}\otimes   \ket{-E_{n}}_{2}$. By following the same procedure, we can finally obtain $\ket{\Psi_{\tau}}=\sum_{n,m}c_n\bramatket{m'}{U_{\tau}}{n}\ket{m'}_{1}\otimes\ket{n}_{R}\otimes\ket{w_{mn}}_{2}\approx\sum_{w}\alpha_w\ket{\phi_w}_{1,R}\otimes\ket{w}_{2}$, where $\ket{\phi_w} =  
1/\alpha_w \sum_{n,m} c_n\,
\bramatket{m'}{U_{\tau}}{n}\,
\delta_{w,w_{mn}} \ket{m'}_{1}\otimes\ket{n}_{R}$ and the term $\ket{m'}_{1}\otimes\ket{n}_{R}\otimes\ket{w_{mn}}_{2}$ describes the encoding of the energy transition path,  $E_n\to E_{m}'$. 
Then, we can obtain Eq.~\eqref{eq:alpha_TTM}. In this case and inn contrast to our previous analysis of the TTM scheme without $\mathcal{H}_{R}$, $\ket{\phi_w}$ becomes orthogonal, i.e. $\braket{\phi_{w'}|\phi_{w}}=\delta_{w,w'}$. However, including an additional subsystem requires to consider the additional information cost. Note that the corresponding cost analysis is beyond the scope of the original work meter model, and we leave this as the focus of future work on the topic.}

\paragraph{Concluding remarks}
\label{sec:conclusion}

In conclusion, employing the work-meter model, we have shown that for finite-dimensional systems the work statistics of the one-time measurement (OTM) scheme follow directly from the Born rule as a natural consequence of envariance for every driving unitary, without involving any explicit measurement postulate, in contrast to the two-time measurement (TTM) scheme, where such an identification is possible only for a Haar-measure-zero set of driving unitaries. In this sense, the OTM scheme is more consistent, provides a clearer conceptual foundation than TTM, and avoids the ambiguities associated with the energetic cost of the measurement process.

We want to note that the other approach to construct the work statistics without taking into account the measurement postulate is based on the geometric structure of quantum state space, as developed by Anza and Crutchfield~\cite{anza2024maximum,anza2021beyond,anza2022geometric,anza2022quantum}. In this framework, ensembles are treated as probability measures over the manifold of pure states, giving rise to ``geometric quantum states'' that encode the fine-grained structure of quantum ensembles beyond the density matrix. This allows for an intrinsic, measurement-independent definition of work. 

While the geometric approach provides a measurement-independent framework, the OTM scheme naturally yields the work statistics directly from the fundamental symmetry of entanglement. This allows for a conceptually transparent and operationally grounded definition of work without introducing additional mathematical structures.

\acknowledgements{We thank Wojciech H. Zurek for insightful discussions and many years of mentorship. A.S and E.D acknowledges U.S. NSF under Grant No. OSI-2328774. A.S also acknowledges PHY-2425180 and Cooperative Agreement PHY-2019786.  A.T is supported by the U.S. Department of Energy under the LDRD program at Los Alamos. S.D acknowledges support from the John Templeton Foundation under Grant No. 63626. This work was supported by the U.S. Department of Energy, Office of Basic Energy Sciences, Quantum Information Science program in Chemical Sciences, Geosciences, and Biosciences, under Award No. DE-SC0025997.}

\bibliography{ref.bib}

\appendix
\newpage 

\onecolumngrid

\onecolumngrid

\setcounter{proposition}{0}
\setcounter{figure}{0}
\setcounter{corollary}{0}
\renewcommand{\figurename}{SUP FIG.}

\newcommand{\SMsection}[1]{%
    \refstepcounter{section}%
    \section*{\thesection.\quad #1}%
}

\newcommand{\SMsubsection}[1]{%
    \refstepcounter{subsection}%
    \subsection*{\thesubsection.\quad #1}%
}

\newcommand{\SMsubsubsection}[1]{%
    \refstepcounter{subsubsection}%
    \subsubsection*{\thesubsubsection.\quad #1}%
}

\setcounter{page}{1}
\setcounter{equation}{0}
\renewcommand{\theequation}{S\arabic{equation}}

\section*{\large{Supplementary Information for\\ ``Quantum Work Statistics from Entanglement-Assisted Invariance''} }

\SMsection{{TTM Work Distribution for Haar-Measure-Zero Sets of Unitaries}}

Here we strengthen the statement in the main text and show that the TTM work states are mutually orthogonal only for a Haar-measure-zero set of driving unitaries. For every work value $w$, define the unnormalized branch state
\begin{align}
\ket{A_w(U_{\tau})}
\equiv
\sum_{m,n}c_n\bramatket{m'}{U_{\tau}}{n}
\delta_{w,w_{mn}}\ket{m'}\,.
\label{SIeq:unnormalized_TTM_branch}
\end{align}
Thus, whenever $\alpha_w\neq 0$, $\ket{\phi_w}$ in the main text is $\ket{\phi_w}=\ket{A_w}/\alpha_w$. For two distinct work values $w\neq w'$, introduce the unnormalized overlap
\begin{align}
G_{w,w'}(U_{\tau})
&\equiv \braket{A_w(U_{\tau})|A_{w'}(U_{\tau})}
\nonumber\\
&=\sum_{m,n,k}c_n^*c_{k}
\bramatket{n}{U_{\tau}^{\dagger}}{m'}
\bramatket{m'}{U_{\tau}}{k}
\delta_{w,w_{mn}}\delta_{w',w_{mk}}\,.
\label{SIeq:G_TTM}
\end{align}
The terms with $n=k$ vanish when $w\neq w'$. Moreover, $G_{w,w'}=\alpha_w^*\alpha_{w'}\braket{\phi_w|\phi_{w'}}$ whenever both branch weights are nonzero. Therefore, the TTM branch states with nonzero weights are mutually orthogonal if and only if
\begin{align}
U_{\tau}\in\mathcal{Z}
\equiv
\left\{
U\in\mathrm{U}(d):
G_{w,w'}(U)=0\text{ for every }w\neq w'
\right\}\,.
\label{SIeq:exceptional_set}
\end{align}

\begin{theorem}
\label{thm:TTM_measure_zero}
Assume that $d\geq 2$, that the initial state has support on at least two distinct initial energies, i.e., there exist $p\neq q$ such that
\begin{align}
c_p\neq 0,\qquad c_q\neq 0,\qquad E_p\neq E_q\,,
\end{align}
and that the final Hamiltonian has at least two distinct energy levels, i.e., there exist $r\neq s$ such that $E_r'\neq E_s'$. Then
\begin{align}
\mu_{\mathrm{Haar}}(\mathcal{Z})=0\,.
\end{align}
Equivalently, for a Haar-random driving unitary, the TTM branch states fail to be mutually orthogonal with probability one.
\end{theorem}

\begin{proof}
Choose indices $p,q,r,s$ satisfying the assumptions and define
\begin{align}
w=E_r'-E_p,\qquad w'=E_r'-E_q\,.
\end{align}
Since $E_p\neq E_q$, one has $w\neq w'$. Equation~\eqref{SIeq:G_TTM} shows that
\begin{align}
F_{w,w'}(U)\equiv\abs{G_{w,w'}(U)}^2
\end{align}
is a real analytic function of the real and imaginary parts of the matrix elements of $U$. A nontrivial real analytic function on a connected open subset of Euclidean space has a zero set of measure zero~\cite{Mityagin2020Zero}. Applying this result in a finite analytic atlas of connected coordinate charts on the compact manifold $\mathrm{U}(d)$ gives the corresponding statement on $\mathrm{U}(d)$: connectedness and analytic continuation rule out an identically vanishing chart restriction when the function is not identically zero, and Haar measure has a smooth positive local density. It therefore remains only to show that $F_{w,w'}$ is not identically zero.

Choose a bijection
\begin{align}
\sigma:\{1,\ldots,d\}\setminus\{p,q\}
\longrightarrow
\{1,\ldots,d\}\setminus\{r,s\}
\end{align}
and, for $0<\theta<\pi/2$, define
\begin{align}
U_{\theta}\ket{p}
&=\cos\theta\ket{r'}-\sin\theta\ket{s'}\,,\\
U_{\theta}\ket{q}
&=\sin\theta\ket{r'}+\cos\theta\ket{s'}\,,\\
U_{\theta}\ket{n}
&=\ket{\sigma(n)'}\,,\qquad n\notin\{p,q\}\,.
\end{align}
The images of the initial energy basis form an orthonormal basis, and hence $U_{\theta}\in\mathrm{U}(d)$. In row $r$, the ordered pair $(p,q)$ contributes
\begin{align}
c_p^*c_q\cos\theta\sin\theta
\end{align}
to $G_{w,w'}(U_{\theta})$. No other ordered pair in that row contributes. Row $s$ cannot contribute the same ordered pair because this would require $E_s'=E_r'$. A reversed contribution $(q,p)$ from row $s$ would require both
\begin{align}
E_s'-E_r'=E_q-E_p,
\qquad
E_s'-E_r'=E_p-E_q\,,
\end{align}
which would imply $E_p=E_q$. Every other row has only one nonzero entry and therefore cannot contribute a product with $n\neq k$. Consequently,
\begin{align}
G_{w,w'}(U_{\theta})
=c_p^*c_q\cos\theta\sin\theta\neq 0\,.
\end{align}
Thus, $F_{w,w'}$ is not identically zero, and
\begin{align}
\mu_{\mathrm{Haar}}
\left(
\left\{U:G_{w,w'}(U)=0\right\}
\right)=0\,.
\end{align}
Since $\mathcal{Z}$ is a subset of this zero set, Eq.~\eqref{SIeq:exceptional_set} has Haar measure zero.
\end{proof}

The measure-zero statement already establishes that the failure of TTM orthogonality is (Haar) generic. For a more general statement, we can also derive a concentration-of-measure estimate showing that other non-Haar random unitaries also lead to the failure of TTM orthogonality with a probability arbitrarily close to one. In particular, a concentration-of-measure estimate gives a quantitative robustness statement. Fix the pair $w,w'$ used in the proof of Theorem~\ref{thm:TTM_measure_zero} and define
\begin{align}
g(U)\equiv\abs{G_{w,w'}(U)},
\qquad
\overline{g}\equiv\mathbb{E}_{\mathrm{Haar}}[g(U)]\,.
\end{align}
The proof above shows that $g(U)>0$ for almost every $U$, and hence $\overline{g}>0$.

To establish the required Lipschitz property without assuming nondegenerate Hamiltonians, let $\Pi_{\varepsilon}$ and $\Pi_{\varepsilon'}'$ denote the spectral projectors of $H_0$ and $H_{\tau}$, respectively. 
Equation~\eqref{SIeq:unnormalized_TTM_branch} can be written as
\begin{align}
\ket{A_w(U)}
=\sum_{\substack{\varepsilon', \varepsilon;\\ \varepsilon'-\varepsilon =w}}
\Pi_{\varepsilon'}'U\Pi_{\varepsilon}\ket{\psi}\,.
\end{align}
The orthogonality of distinct spectral projectors gives
\begin{align}
\norm{A_w(U)}^2
&=\sum_{\substack{\varepsilon', \varepsilon;\\ \varepsilon'-\varepsilon =w}}
\norm{\Pi_{\varepsilon'}'U\Pi_{\varepsilon}\ket{\psi}}^2
\leq 1\,,
\label{SIeq:Aw_norm_bound}\\
\norm{A_w(U)-A_w(V)}^2
&=\sum_{\substack{\varepsilon', \varepsilon;\\ \varepsilon'-\varepsilon =w}}
\norm{\Pi_{\varepsilon'}'(U-V)\Pi_{\varepsilon}\ket{\psi}}^2
\nonumber\\
&\leq \norm{U-V}_{\mathrm{op}}^2
\sum_{\varepsilon}\norm{\Pi_{\varepsilon}\ket{\psi}}^2
\leq \norm{U-V}_{\mathrm{HS}}^2\,.
\label{SIeq:Aw_Lipschitz}
\end{align}
It follows from Eqs.~\eqref{SIeq:Aw_norm_bound}-\eqref{SIeq:Aw_Lipschitz} and the reverse triangle and Cauchy--Schwarz inequalities that
\begin{align}
\begin{split}
\abs{g(U)-g(V)}
&\leq \abs{G_{w,w'}(U)-G_{w,w'}(V)}\\
&\leq
\norm{A_w(U)-A_w(V)}\norm{A_{w'}(U)} + \norm{A_{w'}(U)-A_{w'}(V)}
\norm{A_w(V)}\\
&\leq 2\norm{U-V}_{\mathrm{HS}}\,.
\end{split}
\label{SIeq:g_Lipschitz}
\end{align}
Thus, $g$ is $2$-Lipschitz with respect to the Hilbert--Schmidt metric. The logarithmic-Sobolev concentration inequality for Haar measure on $\mathrm{U}(d)$ (stated as Corollary~17 of Ref.~\cite{MeckesMeckes2013Spectral}) applied to $-g$, then yields, for every $0<t<\overline{g}$,
\begin{align}
\mu_{\mathrm{Haar}}
\left(
 g(U)\leq \overline{g}-t
\right)
\leq
\exp\left(-\frac{d\,t^2}{48}\right)\,.
\label{SIeq:TTM_concentration}
\end{align}
In particular,
\begin{align}
\mu_{\mathrm{Haar}}
\left(
 g(U)\leq \alpha \overline{g}
\right)
\leq
\exp\left(-\frac{d (1-\alpha)^{\,2} \,\overline{g}^{\,2}}{48}\right)\,,
\label{SIeq:TTM_concentration_half_mean}
\end{align}
with $0 \leq \alpha <1 $. Therefore, a Haar-random unitary violates exact TTM orthogonality with almost probability one, and the probability that this overlap witness lies substantially below its Haar mean is given by Eq.~\eqref{SIeq:TTM_concentration}.

\SMsection{{TTM Work Distribution as Born Probability with Additional Reference System}}

Here, we demonstrate that the TTM work distribution can also be derived as Born probability distribution, by adding a reference subsystem $\mathcal{H}_{R}$ that encodes the eigenstates of the initial Hamiltonian into its displacement. 

Let $\Gamma_A$ describe a controlled translation generated by a Hermitian operator on $\mathcal{H}_{R}$ and $\mathcal{H}_{2}$ as 
$\Gamma_A$, 
\begin{align}
    \Gamma_A = \sum_{n}\dya{n}\otimes e^{-i n P}\otimes e^{iE_{n} P}\,,
\end{align}
Then, we can write 
\begin{align}
\ket{\psi_0} &= \Gamma_A\ket{\psi_{<0}}= \sum_{n} c_n 
\ket{n}_{1}   \otimes\ket{n}_{R}\otimes   \ket{-E_{n}}_{2}\,.
\end{align}
and the initial reduced state of $\mathcal{H}_1$,
\begin{align}
    \rho_0\equiv \Tr_2\left[\dya{\psi_0}\right]=\sum_{n}\abs{c_n}^2\dya{n}\,.
\label{SIeq:initial_{1}tate}
\end{align}
We evolve $\ket{\psi_0}$ from $t=0$ to $t=\tau$ under $U_{t} \otimes\id_{R}\otimes \id_{2}$, we then obtain 
\begin{align}
\ket{\psi_{\tau}}=\sum_{n}c_n U_{\tau}\ket{n}_{1}\otimes\ket{n}_{R} \otimes \ket{-E_{n}}_{2}\,.
\end{align}
Next, we again turn on the system-meter interaction by applying a unitary
\begin{align}
    \Gamma_B = \sum_{m}\dya{m'}\otimes\id_{R}\otimes e^{-i E_{m}' P}\,.
\end{align}
Then, after the discretization, we obtain $\ket{\Psi_{\tau}}=\Gamma_B\ket{\psi_{\tau}}$ as 
\begin{align}
\begin{split} 
\ket{\Psi_{\tau}}
&=\sum_{n,m}c_n\bramatket{m'}{U_{\tau}}{n}\ket{m'}_{1}\otimes\ket{n}_{R}\otimes\ket{w_{mn}}_{2}\\
&\approx\sum_{w}\left(\sum_{n,m}c_n\bramatket{m'}{U_{\tau}}{n}\delta_{w,w_{mn}}\ket{m'}_{1}\otimes\ket{n}_{R}\right)\otimes\ket{w}_{2}\\
&=\sum_{w}\alpha_w\ket{\phi_w}\otimes\ket{w}_{2}\,,
\end{split}
\label{SIeq:TTM_env}
\end{align}
where again $w_{mn}\equiv E_{m}'-E_n$ and $\braket{w|w'}=\delta_{w,w'}$.  Here, for this scenario, $\ket{\phi_w}$ is given as 
\begin{align}
\ket{\phi_w}\equiv 
\frac{1}{\alpha_w} \sum_{n,m} c_n\,
\bramatket{m'}{U_{\tau}}{n}\,
\delta_{w,w_{mn}} \ket{m'}_{1}\otimes\ket{n}_{R}\,.
\label{SIeq:phi_wTTM}
\end{align}
Because $\ket{\phi_w}$ is a pure state, using the orthonormal condition 
$\braket{k|\ell}=\delta_{k,\ell}=\braket{k'|\ell'}$ and
$\delta_{w,w_{mn}}^2=\delta_{w,w_{mn}}$, we can obtain
\begin{align}
  \abs{\alpha_w}^2
  = \sum_{n,m} \abs{c_n}^2
\abs{\bramatket{m'}{U_{\tau}}{n}}^2
    \delta_{w,w_{mn}} \,,
\end{align}
which again coincides with the TTM work distribution. Actually, in this case, $\ket{\phi_w}$ becomes orthogonal as 
\begin{align}
\begin{split}
\braket{\phi_{w'}|\phi_{w}}&=\frac{1}{\alpha_{w'}^{*}\alpha_{w}}\sum_{\substack{n,m\\ k,\ell}}c_{k}^{*}c_{n}\bramatket{k}{U_{\tau}\ad}{\ell'}\bramatket{m'}{U_{\tau}}{n}\delta_{w',w_{\ell k}}\delta_{w,w_{mn}}\braket{\ell'|m'}\braket{k|n}\\
&=\frac{1}{\alpha_{w'}^{*}\alpha_{w}}\sum_{\substack{n,m\\ k,\ell}}c_{k}^{*}c_{n}\bramatket{k}{U_{\tau}\ad}{\ell'}\bramatket{m'}{U_{\tau}}{n}\delta_{w',w_{\ell k}}\delta_{w,w_{mn}}\delta_{\ell,m}\delta_{k,n}\\
&=\frac{1}{\alpha_{w'}^{*}\alpha_{w}}\sum_{n,m}\abs{c_n}^2\abs{\bramatket{m'}{U_{\tau}}{n}}^2\delta_{w',w_{mn}}\delta_{w,w_{mn}}\\
&=\delta_{w,w'}\,.
\end{split}
\end{align}
Therefore, similar to the OTM scheme in the main text, we can now derive the work statistics from the entanglement with the work meter given that a reference system is utilized to record the initial energy eigenstates.

\end{document}